# A 3-D Full-Wave Model to Study the Impact of Soybean Components and Structure on L-Band Backscatter

Kaiser Niknam, *Member, IEEE*, Jasmeet Judge, *Senior Member, IEEE*, A. Kaleo Roberts, *Student Member, IEEE*,

Alejandro Monsivais-Huertero, *Senior Member, IEEE*, Robert Moore, *Senior Member, IEEE*,

Kamal Sarabandi, *Life Fellow, IEEE*, and Jiayi Wu, *Student Member, IEEE*

*Abstract*—**Microwave remote sensing offers a powerful tool for monitoring the growth of short, dense vegetation like soybean. As the plants mature, changes in their biomass and 3-D structure impact the electromagnetic (EM) backscatter signal. This backscatter information holds valuable insights into crop health and yield, prompting the need for a comprehensive understanding of how structural and biophysical properties of soybeans as well as soil characteristics contribute to the overall backscatter signature. In this study, a full-wave model is developed for simulating L-band backscatter from soybean fields. Leveraging the ANSYS High-Frequency Structure Simulator (HFSS) framework, the model solves for the scattering of EM waves from realistic 3-D structural models of soybean, explicitly incorporating the interplant scattering effects. The model estimates of backscatter match well with the field observations from the SMAPVEX16-MicroWEX and SMAPVEX12, with average differences of 1-2 dB for co-pol and less than 4 dB for cross-pol. Furthermore, the model effectively replicates the temporal dynamics of crop backscatter throughout the growing season.**

Manuscript received XX xx, 2023; revised XX xx, 2024; and accepted XX xx, 2024. This work is supported by NASA Grant #80NSSC21K0186 (corresponding author: Jasmeet Judge).

Kaiser Niknam, Jasmeet Judge, and Jiayi Wu are with the Center for Remote Sensing, Department of Agricultural and Biological Engineering, Institute of Food and Agricultural Sciences, University of Florida, Gainesville, Florida, USA (e-mail: {gheysar.niknam; jasmeet; wuj2}@ufl.edu).

A. Kaleo Roberts and Kamal Sarabandi are with the Department of Electrical and Computer Engineering, University of Michigan, Ann Arbor, Michigan, USA (e-mail: {kaleor; saraband}@umich.edu).

Alejandro Monsiváis-Huertero is with the Escuela Superior de Ingenieria Mecanica y Electrica, Ticoman, Instituto Politecnico Nacional, Mexico City, Mexico (e-mail: monsivais@ufl.edu).

Robert C. Moore is with the Department of Electrical and Computer Engineering, University of Florida, Gainesville, Florida, USA (e-mail: moore@ece.ufl.edu).

The HFSS analysis revealed that the stems and pods are the primary contributors to HH-pol backscatter, while the branches contribute to VV-pol, and leaves impact the cross-pol signatures. In addition, a sensitivity study with 3-D bare soil surface resulted in an average variation of 8 dB in co- and cross-pol, even when the root mean square height and correlation length were held constant. These capabilities underscore the model's potential to provide insights into the underlying dynamics of the backscatter for growing vegetation.

## I. INTRODUCTION

MICROWAVE monitoring of soybean fields is an efficient method to assess soybean plant growth over the course of a growing season [1]. The dielectric properties and 3-D structure of the soybean plant change as it grows, and these changes impact the backscatter from the crop in meaningful ways [2] [3] [4]. As a result, radar data provides valuable information on crop health and yield [5] [6] [7]. Thus, an understanding of how structural and biophysical properties of soybean and soil characteristics contribute to the overall signature during the growing season is required to decode the information encoded in radar backscatter. Empirical models have been used to estimate backscatter from crops using relationships that relate vegetation water content (VWC) and soil properties (moisture and roughness) to backscatter values [8] [9] [10] [11]. However, these relationships are specific to the region that the model was calibrated for. Additionally, understanding the mechanisms underlying crop backscatter is best achieved by forward models that simulate the interaction of electromagnetic (EM) waves with crops. The development of more accurate reverse models is critically dependent on the insights gained from such forward modeling approaches.

Modeling the interaction between EM waves and growing vegetation in agricultural fields is a challenge due to its structural complexity. The water cloud model (WCM) [12] [13] [14] [15] [16], radiative transfer equation (RTE) [17] [18] [19], and distorted Born approximation (DBA) [20] [21] [22] [23] [24] [25] have been methods of choice to model backscatter from vegetation canopies. These methods are based on the simplified models of the plant and/or simplified assumptions on how EM waves interact with the crop. For example, in WCM, canopies are modeled as a water cloud containing identical water droplets uniformly distributed within the canopy [26]. Additionally, scattering from the crop is reduced to the scattering from its canopy. In DBA models, however more realistic compared to WCM, scattering mechanisms at the L-band frequencies (1-2 GHz) are still limited to [27] direct scatter (e.g., direct scattering from ground, or, direct scattering from vegetation), double-bounce scatter (e.g., ground-vegetation scattering, or, ground-vegetation-ground scattering), and volume scattering (e.g., attenuating within, and scattering from canopy). Although these simplifying assumptions helped to implement a set of successful early models [10] [21] [13] [20], they did not work well in all conditions. In particular, they fail (1) when the plants and/or their canopies are not uniformly distributed, (2) when the scattering from the plant cannot be decomposed into a summation of independent scatterings from its components, and (3) when far-field assumption held throughout the RTE and DBA models does not hold true [28] [29].

The above-mentioned methods have a common aspect, namely, they are computationally efficient. Recent advances in computational resources have motivated the development of 3-D models based on numerical techniques such as Numerical Maxwell Model in 3-D (NMM3D) [30], mainly accompanied by hybrid methods [29] [31]. In hybrid methods, Maxwell's equations are solved for isolated realistic 3-D structural models of the plant to impose less computational cost, and then transition matrices (T-matrices) obtained from these solutions are employed to build canopies and simulate the total backscatter from a field. Although hybrid approaches are computationally efficient, they tend to ignore physical overlaps and multiple scattering existing in the crops, and could create errors of several orders of magnitude (up to 20 dB) in Radar Cross Section (RCS) at lower incidence angles [32] [33].

To address the limitations of hybrid methods, full-wave simulations have recently been used to estimate backscattering from fields without making any restrictive assumptions about the interaction of EM waves with crops [34] [35] [36] [37]. These models are computationally more intensive than previous methods, but they provide a robust framework to model backscatter from crops under different conditions. Recently, Roberts *et al.* [38] have developed a full-wave simulation model that estimates corn backscatter using a 3-D structural mesh for a complete plant. This approach improves upon hybrid methods by considering multiple scatterings within crop canopies, and upon previous full-wave models by employing a realistic 3-D structural plant model for the growing crop. Roberts *et al.* [38] evaluated the model for corn, however, such a model can be adapted for different corps and growth stages. In this study, the model was modified and extended to investigate backscatter from growing soybean, which is a structurally different crop, lacking a dominant stem, and facing additional computation challenges from overlapping vegetation components.

The remainder of this paper is organized as follows: In Section II, the model structure and simulation setup and the data used for validation are described. In Section III, backscatter estimates from the full-wave model are compared with field observations. Section IV delves into the model's capability to accurately estimate the overall signature throughout the growing season, assess the impact of soil roughness on backscatter estimates, and investigate the contributions of various vegetation components to the overall signature. Lastly, Section V summarizes the novel contributions and findings of the study.

## II. Methods

### A. HFSS Modeling

ANSYS High-Frequency Structure Simulator (HFSS) was the EM solver used for this study, which presents several advantages including high accuracy of solutions using the finite element method (FEM) and a user-friendly graphical interface that allows users to build or import 3-D structural models. In this study, we used a unit cell with periodic boundary conditions that replicate the unit cell infinitely in all directions, simulating an infinite field. HFSS generated and solved a single mesh for the unit cell, significantly reducing memory usage and computational costs.

As shown in Fig. 1a, a typical unit cell included soybean crops planted in rows along the y-direction at an average inter-plant distance of $\Delta y$. The rows were considered at an average row distance of $\Delta x$ along the x-direction from each other, and $\Delta x/2$ from the top and bottom borders of the unit cell. Similarly, the soybean plants were located at a distance of $\Delta y$ from each other, and $\Delta y/2$ from the left and right boundaries of the unit cell (Fig. 1b). As a result, the width and length of unit cells were integer multiples of the average distances between the soybean plants along and across rows, respectively. Here, unit cells are described by two integers as $n \times m$ where $n$ is the number of rows, and $m$ is the number of soybeans in each row. Unit cells were coupled with periodic boundary conditions on sides, and terminated by a perfectly matched layer (PML) on top, and a layer of soil with a rough surface, and then, a corresponding surface impedance boundary condition (SIBC) on the bottom (Fig. 1a). Periodic boundary conditions require that the EM wave's path inside the computational grid matches the wavelength ($\lambda$) of the incident wave, causing the elevation (φ) and azimuth (θ) angles of incidence follow (1) and (2): [38]

$$\varphi = \tan^{-1}\left[\left(L_x/p\right)/\left(L_y/q\right)\right] \qquad (1)$$

$$\theta = \sin^{-1}\left[\frac{\lambda}{2}\sqrt{\left(p/L_x\right)^2 + \left(q/L_y\right)^2}\right] \qquad (2)$$

where $L_x$ and $L_y$ are the width and length of the unit cell, respectively, and $p$ and $q$ are arbitrary integers. φ and θ hold in (1) and (2) are called Bragg angles. In this study, the simulations were conducted at incidence angles close to Bragg angles. Unit cells were uniformly illuminated by a plane wave at given incidence angles and frequency.

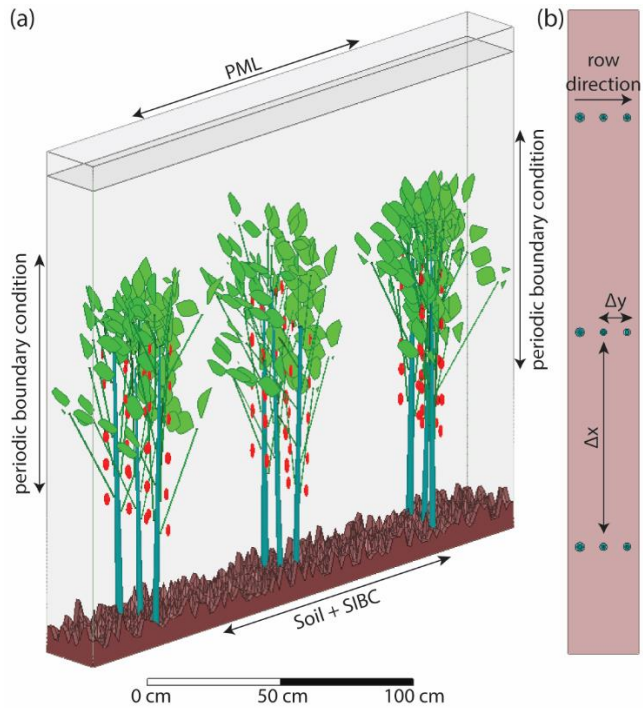

**Fig. 1. A typical HFSS unit cell.** (a) The side view of a unit cell employed in this study is shown here. PML stands for perfectly matched layer, and SIBC stands for surface impedance boundary condition. (b) The top view of the unit cell is shown. $\Delta x$ and $\Delta y$ are the distances between soybean plants across, and along rows, respectively.

Inputs to HFSS included soil roughness, dielectric properties of soil and vegetation components, and 3-D plant structure. The soil roughness was defined by the correlation length and root mean square (RMS) height [39]. The complex permittivity of soil was estimated using the Mironov model [40] as a function of volumetric soil moisture (VSM), which is the ratio of the volume of water in a soil sample to the total volume of the sample, as well as frequency and clay percent. The complex permittivities of stems, branches, leaves, and pods were estimated based on the gravimetric moisture content ($M_g$) and physical temperature [41] [42]. $M_g$ is the ratio of the mass of water in a component to the total mass of the component. All simulations were performed at L-band (1.26 GHz).

The 3-D structural models of growing soybean plants consisted of stems, branches, leaves, and pods (during the reproductive stages). Stems and branches were frustums of cones described by top and base radii, length, and azimuth and elevation angles of rotation around their major axes. Leaves were 2-D impedance sheets with widths and lengths but no thicknesses, following [43] and [38]. No significant difference was found between the backscatter coefficients ($\sigma^0$) using 2-D and 3-D leaves, however, 2-D leaves significantly reduced the computational time and memory requirements. The branches were positioned at node heights, with leaves positioned at the end of the branches as trifoliates. Leaves were rotated around branches in azimuth and elevation directions. Finally, each cluster of pods was represented by one 3-D ellipsoid with the same volume as the cluster, described by the lengths of three semi-axes. The clusters of pods were placed at a fixed distance of 2 cm below the base of the branches with their major axis perpendicular to the ground surface. Since the backscatter contribution of the thin stalks that connect leaves to branches, called petioles, was found to be minimal, they were eliminated to reduce computational complexity.

In this study, the parameters for generating the 3-D structural models for soybean plants were obtained from field experiments, described in Section II.B. The variabilities observed in the field data were used to generate multiple realizations with perturbed branch and stem angles as well as plants' locations within the unit cell. Twenty realizations were found to provide a confidence interval of ±2.5 dB, similar to the measurement fidelity in the experiments.

*B. Field Experiments*

The study used inputs and observations from the Soil Moisture Active Passive Validation Experiment (SMAPVEX) 2016 – Microwave, Water, and Energy Balance Experiment (MicroWEX), and SMAPVEX in 2012 (SMAPVEX12).

*B-1) SMAPVEX16-MicroWEX*

Measurements from SMAPVEX16-MicroWEX [44] were used to provide inputs to the model and evaluate its performance. The experiment was conducted over the entire growing season of soybean and corn from May 23 (DoY 144) to September 2 (DoY 246), 2016, on a commercial farm (Sweeney Farms) in Alden, Iowa (Fig. 2a). The goal of the experiment was to collect high temporal resolution active and passive microwave data from soybean and corn fields to validate and improve the SMAP retrieval measurements of soil moisture and vegetation water content. It involved ground-based measurements from soybean and corn fields at L-band (1.41 GHz for passive and 1.26 GHz for active) [44]. The $\sigma^0$ values were obtained hourly at azimuth angles of $-9°$, $0°$, and $9°$, and an elevation angle of $40°$, with respect to the direction perpendicular to row structure. At the soybean field, the observations were performed over two periods, from July 8-12 (DoY 190-194) during the early growing season, and September 2-6 (DoY 246-250) during the late/reproductive stage. This study used observations at 6 am, matching anticipated satellite observations such as those from NASA-ISRO Synthetic Aperture Radar (NISAR).

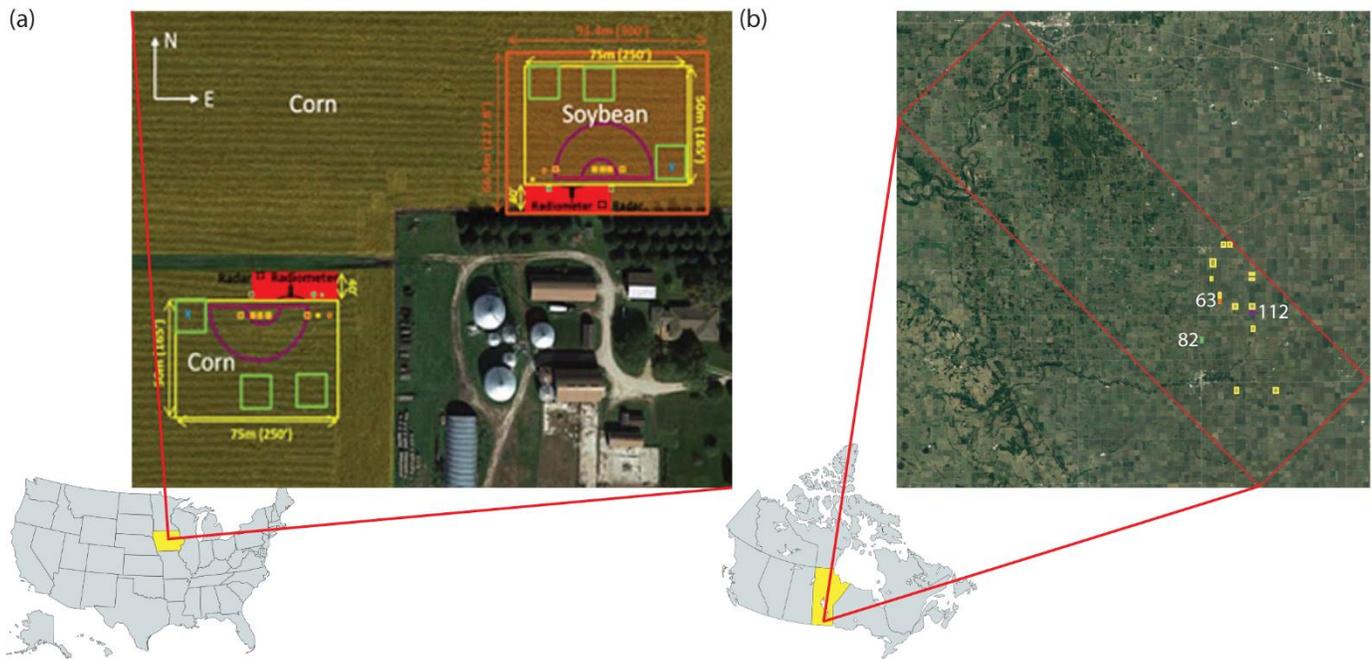

**Fig. 2. Study areas**. (a) The SMAPVEX16-MicroWEX study area located in Central Iowa (inset) and its location relative to the US map are displayed here. (b) The SMAPVEX12 study area (inset) located in Manitoba, Canada, and its location relative to the Canada map are displayed here. The yellow rectangles depict the soybean fields surveyed during SMAPVEX12. The selected fields #63, #82, and #112 are designated by orange, green, and purple rectangles, respectively.

In addition to the microwave observations, SMAPVEX16-MicroWEX involved weekly vegetation measurements of row spacing; plant density; soil moisture, texture, and roughness; and geometry and wet and dry biomass of the vegetation components of the plant at three sampling sites [44]. The geometrical features measured during the experiment included heights and diameters of stems and branches, heights of branch-nodes, sizes of leaves and pods, and the number of pods and leaves per branch. The vegetation sampling conducted on 11 days: June 22 (DoY 174), June 28 (DoY 180), July 5 (DoY 187), July 11 (DoY 193), July 18 (DoY 200), July 26 (DoY 208), August 1 (DoY 214), August 7 (DoY 220), August 18 (DoY 231), August 236 (DoY 236), and September 1 (DoY 245). The vegetation samples were weighed wet and dried at 48°C for 3-7 days to obtain $M_g$. Soil moisture was recorded every 15 minutes at depths of 2, 5, 10, 15, 30, 60, and 120 cm. The average of soil moisture at depths 2 and 5 cm was considered as the representative soil moisture for this study. Soil roughness was measured using a pin profiler on two days, May 25 (DoY 146), and August 10 (DoY 223).

*B-2) SMAPVEX12*

SMAPVEX12 was conducted over a short, 5-week, non-reproductive period of the growing season, from June 7 - July 19

(DoY 159-201), 2012, covering more than 15 soybean and 10 corn fields near Manitoba, Canada (Fig. 2b). During the experiment, airborne observations of $\sigma^0$ were obtained by NASA's Unmanned Aerial Vehicle Synthetic Aperture Radar, UAVSAR. Fully polarimetric $\sigma^0$ values at L-band (1.26 GHz) were recorded at an azimuth angle of 45°, and at elevation angles of 30° to 50° (30° to 60° for cross-polarization) [45]. However, $\sigma^0$ values were corrected to 40° elevation angle based on the histogram-based (HIST) technique [46]. In this study, $\sigma^0$ observations from three soybean fields with minimal variability in $\sigma^0$ measured across the field were chosen for implementation. Since a UAVSAR footprint is about 6 m, the three selected fields with identifiers 62, 82, and 112 contained 10027, 7109, and 9400 spatially distributed SAR measurement pixels, respectively.

During the experiment, ground-based measurements of soil and vegetation parameters were obtained [45] in addition to the field's crop type, geolocation, row direction, and crop density [47]. Soil moisture was measured in the top 6 cm every 1-5 days using theta and hydra probes. Soil texture and roughness were measured once in each of the three fields. Destructive sampling was used to collect vegetation data from three measurement sites in each of the three fields. For soybeans, 10 plants, five each from two rows, were collected at each measurement site, and lengths and diameters of 10 stems were recorded. $M_g$ of crop components were calculated similar to that as in SMAPVEX16-MicroWEX. A detailed description of SMAPVEX12 is provided in [48].

*C. HFSS Simulations*

*C-1) SMAPVEX16-MicroWEX*

Simulations were conducted for the days on which vegetation data were collected, and also for the days during the two periods of $\sigma^0$ observations (Section II.B-1).

The geometrical measurements of six sample soybean plants were used to build 3-D structural models of soybean. The geometrical features such as branch and stem angles that were not observed in the SMAPVEX16-MicroWEX dataset were set based on literature-derived values on typical soybean phenology as follows. The plant spacing within the rows was randomly perturbed by ±2 cm, and the row spacing was perturbed by ±3 cm (Fig. 1b). Stems were rotated randomly around the z- and y-directions, and rotation angles were selected from a Gaussian random distribution with a zero mean and a standard deviation (STD) of 3° for the early days (DoY < 231), and 2° for the late days (to maintain soybean stems inside the space of unit cells). The azimuth angles of branches were obtained from a Gaussian random distribution with a zero mean and an STD of ~3°, based on our observations from other soybean fields. The elevation angles of branches were taken from an exponential random distribution with a mean and an STD of 18° and 1°, respectively, based on our observations from other fields. The elevation angles higher than 45° were discarded.

3-D structural models for DoY 193 and 245 were used for simulations during the two periods with $\sigma^0$ observations, on DoYs 190-194 and 246-250, respectively. $M_g$ values were obtained by interpolating the values measured on the two sampling days. Fig. 3a displays an example of 3-D structural models during the growing season. Figs. 3b-e represent the evolution of the lengths and radii of stems and branches, and lengths and widths of leaves and pods, respectively.

The soil layer (Fig. 1a) thickness of 6 cm was selected to avoid any subsurface reflections with the implemented SIBC. A static soil roughness with an RMS height of 2.59 cm and a correlation length of 34 cm was applied for all simulations based on one of the reported values. Throughout the growing season, the height of unit cells was constant at 180 cm, the maximum height observed, so that the plants were completely inside the unit cell. Based on soybean plant spacing of 8.33 cm in-row and 76 cm across-rows, a 3x3 unit cell with corresponding Bragg angles of $\varphi = 0°$ and $\theta = 39.13°$ was selected to approximate target incidence angles of $\varphi = 0°$ and $\theta = 40°$.

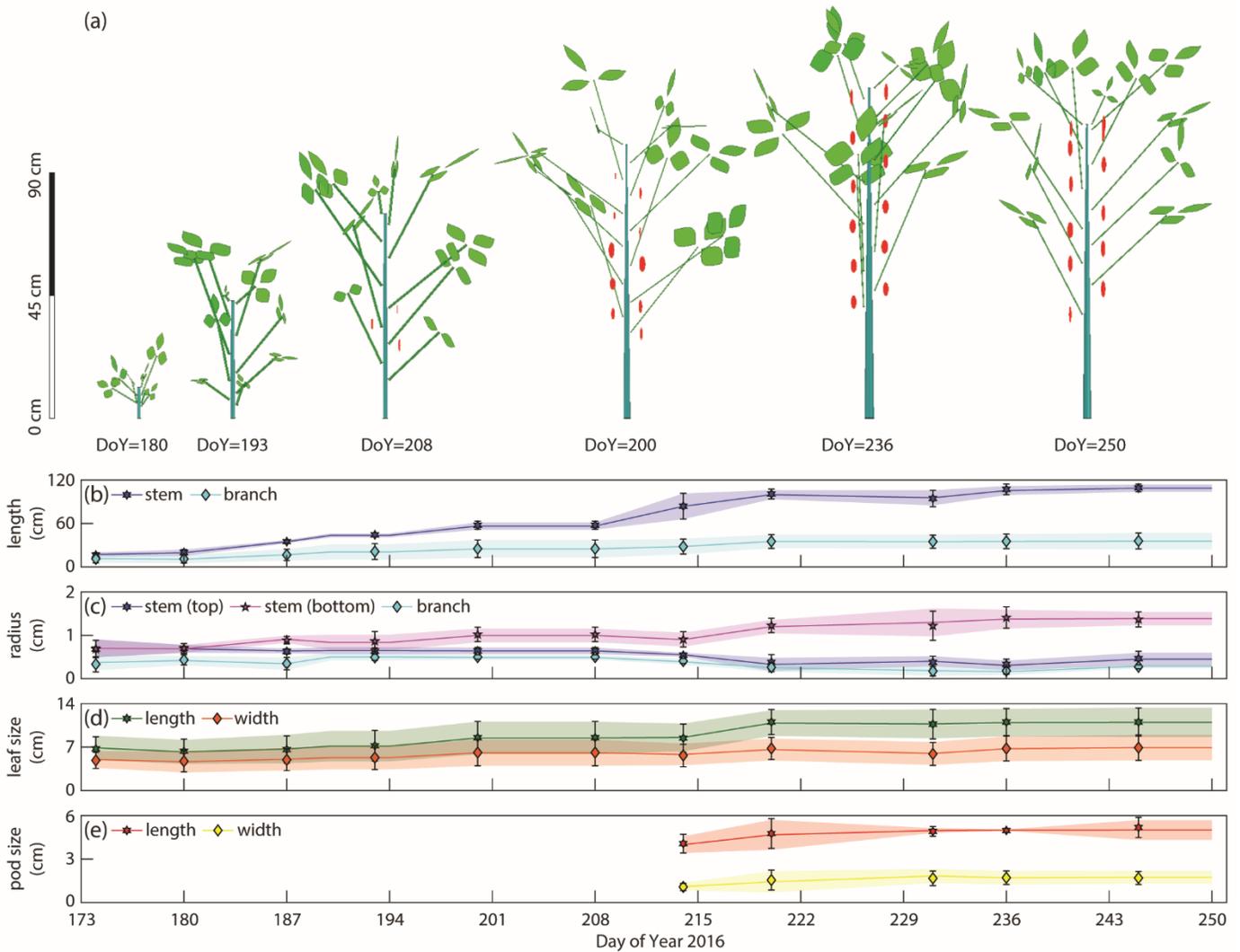

**Fig. 3. The generated 3-D structural models of soybean for the simulations corresponding to SMAPVEX16-MicroWEX.** (a) Sample 3-D structural models are displayed here. The day for which the model is generated is indicated below the model. (b) The evolution of the stems' and branches' length over days is displayed here. The shaded area shows the standard deviation (STD) of the corresponding trace. The error bars display the sampling values and the corresponding STD. (c) The same as panel (b) but for radii of stems and branches (d) The same as panel (b) but for lengths and widths of leaves (e) The same as panel (b) but for lengths and widths of pods.

Since the unit cells were excited by a uniform plane wave source, it approximates the airborne measurements well. However, it is not well suited for ground measurements performed in SMAPVEX16-MicroWEX, resulting in non-uniform illumination weighted by the antenna pattern. To address this problem, a scaling technique used in [49] was employed to translate $\sigma^0$ from the HFSS simulations to $\sigma^0$ from ground-based measurements. When taking angle-dependence into account, the estimated $\sigma^0$ can be given by

$$\sigma^0 = \frac{r_0^4}{A} \int_A \frac{F_t(\varphi,\vartheta).F_r(\varphi,\vartheta).\sigma^0(\varphi,\vartheta)}{r(\varphi,\vartheta)^4} dA, \qquad (3)$$

where $F_t$ and $F_r$ are the normalized beam patterns, $A$ is the antenna footprint, $r_0$ is the distance from the radar to the center of the footprint, and $r$ is the distance from the radar to a scattering center inside $A$. The scaling technique discretizes the integral by dividing $A$ into small rectangles, or tiles, defined by the unit cell's dimensions (Fig. 4), and the $\sigma^0(\varphi, \theta)$ for each tile is obtained from HFSS. Thus the discretized version becomes a coherent summation:

$$\begin{bmatrix} \sigma_{vp}^0 \\ \sigma_{hp}^0 \end{bmatrix} = \frac{4\pi}{A} \sum_n \left| \left(\frac{r_0}{r_n}\right)^2 e^{-2jkr_n} \mathbf{F}_n \mathbf{S}_n \mathbf{F}_n \begin{bmatrix} e_v \\ e_h \end{bmatrix} \right|^2 \qquad (4)$$

Here, $k$ is the wavenumber, $n$ indexes different tiles, $r_n$ is the distance to a tile, and $[e_v, e_h]^T$ indicates vertical ($[1, 0]^T$) or horizontal ($[0, 1]^T$) transmit polarization. The matrix product of the normalized, complex antenna amplitude $\mathbf{F}$ and scattering matrix $\mathbf{S}$ is given by

$$\mathbf{F}_n \mathbf{S}_n \mathbf{F}_n = \begin{bmatrix} f_{vv} & f_{vh} \\ f_{hv} & f_{hh} \end{bmatrix} \begin{bmatrix} S_{vv} & S_{vh} \\ S_{hv} & S_{hh} \end{bmatrix} \begin{bmatrix} f_{vv} & f_{vh} \\ f_{hv} & f_{hh} \end{bmatrix} \qquad (5)$$

and accounts for the coupling between cross-polarization terms in the antenna and scatterer, assuming a monostatic scenario. Then, for a nominal incidence angle ($\varphi_i, \theta_i$), the estimated average $\sigma^0$ is calculated with Monte-Carlo simulations that randomize the locations of the tiles in $A$ and the scattering matrix used for each tile. Due to a random selection of unit cells, the scaling results are prone to outliers. To address this problem, the scaling technique was used 20 times to generate 20 scaled $\sigma^0$ values.

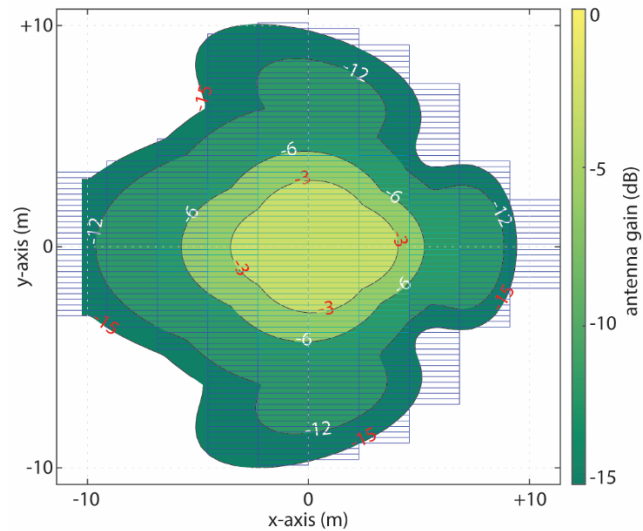

**Fig. 4. The scaling technique to allow HFSS comparisons with ground-based measurements.** The footprint of the University of Florida L-band automated radar system (UFLARS) was tiled by unit cells (blue rectangles), and the received $\sigma^0$ by radar was assumed to be a weighted sum of the $\sigma^0$ values of the tiled unit cells. The heatmap illustrates the weighting applied to each point within the radar footprint.

*C-2) SMAPVEX12*

Simulations were conducted for 12 days during which soil moisture and $\sigma^0$ values were observed: June 17 (DoY 169), June 22 (DoY 174), June 23 (DoY 175), June 25 (DoY 177), June 27 (DoY 179), June 29 (DoY 181), July 5 (DoY 187), July 8 (DoY 190), July 10 (DoY 192), July 13 (DoY 195), July 14 (DoY 196), and July 17 (DoY 199).

3-D structural models were built as follows. The length and diameter of stems were randomly selected from 30 measurements. The parameters such as stem length, diameter, and $M_g$ were interpolated between measurement days, as shown in panels b & c of Figs. 5-7. The top and bottom diameters of the stems were considered the same due to the lack of detailed observations. In addition, 10 branches with the same diameter but 1/3 of the length of stems were designed for each soybean plant. Branches were arranged alternately on either side of the stem, aligned with the positive and negative x-axis directions (shown in Fig. 1b). The length and width of trifoliates were obtained from the SMAPVEX16-MicroWEX dataset for plants with similar heights and widths due to the lack of information during SMAPVEX12.

Similar to the simulations for SMAPVEX16-MicroWEX, the geometrical features that were not observed in the SMAPVEX12 dataset were set randomly. The rotation angles of stems were selected from a Gaussian random distribution with a zero mean and an STD of 1°. The elevation angles of branches were taken from a uniform random distribution between 1° and 60°. The location of branches on the stem were selected from a uniform random distribution between zero and the length of the corresponding stem. Leaves were rotated around the corresponding branch randomly along azimuth and elevation directions. The random values for the rotation angles were drawn from a uniform random distribution between ±60°. The random parameters generated realizations for the simulations.

Similar to SMAPVEX16-MicroWEX simulations, the soil layer had a thickness of 6 cm and its dielectric properties were estimated from the observed VSM. The rough soil was generated based on measured RMS height and correlation length for each field. The same roughness was used for all simulations corresponding to each of the three fields. The height of unit cells was set to 180 cm. The simulations were performed at the measurement frequency of 1.26 GHz.

Fig. 5a displays a sample 3-D structural model generated for five days for field #63. Figs. 5b-d represent the daily evolution of the stem and leaf parameters. The azimuth angles of branches were between ±35° in the early days, but gradually reduced to ±9° in the late days, to keep the branches inside the unit cells. The RMS height and correlation length of the soil roughness were 0.92 and 10 cm, respectively. Based on soybean plant spacing of 5.31 cm in-row and 76.5 cm across-rows, a 2x5 unit cell with corresponding Bragg angles of $\varphi = 43.86°$ and $\theta = 40.32°$ was selected to approximate target incidence angles of $\varphi = 45°$ and $\theta = 40°$.

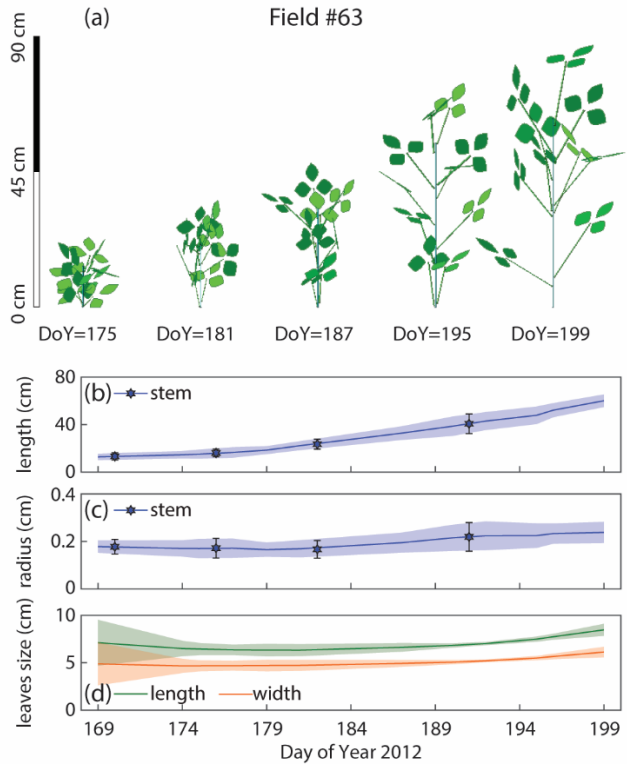

**Fig. 5. The generated 3-D structural models of soybean for field #63.** (a) Sample 3-D structural models generated for 5 days for field #63 are displayed here. (b) The evolution of the stems' length over days is displayed. The shaded area shows the STD. The markers show the measurements, with their STD shown as error bars. (c) The same as panel (b) but for radii of stems. (d) The same as panel (b) but for lengths and widths of leaves.

Fig. 6a displays a sample 3-D structural model generated for five days for field #82. Figs. 6b-d represent the daily evolution of the stem and leaf parameters. The azimuth angles of branches were between ±30° in the early days, but gradually reduced to ±5° in the late days, to keep the branches inside the unit cells. The RMS height and correlation length of the soil roughness were 0.98 and 13 cm, respectively. Based on soybean plant spacing of 4.77 cm in-row and 77.8 cm across-rows, a 2x5 unit cell with corresponding Bragg angles of $\varphi = 47.37°$ and $\theta = 42.63°$ was selected to approximate target incidence angles of $\varphi = 45°$ and $\theta = 40°$.

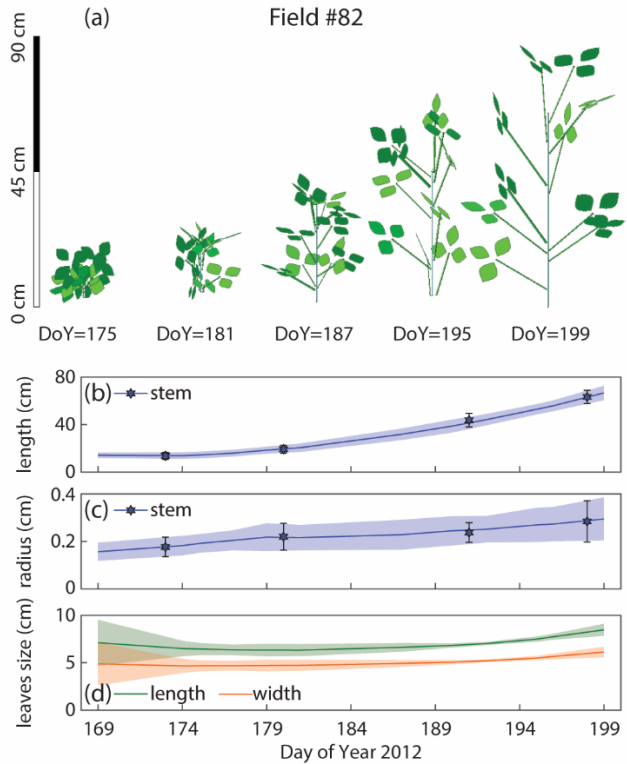

**Fig. 6. The generated 3-D structural models of soybean for field #82.** (a) Sample 3-D structural models generated for 5 days for field #82 are displayed here. (b) The evolution of the stems' length over days is displayed. The shaded area shows the STD. The markers show the measurements, with their STD shown as error bars. (c) The same as panel (b) but for radii of stems. (d) The same as panel (b) but for lengths and widths of leaves.

Fig. 7a displays a sample 3-D structural model generated for five days for field #112. Figs. 7b-d represent the daily evolution of the stem and leaf parameters. The azimuth angles of branches were between ±40° in the early days, but gradually reduced to ±9° in the late days, to keep the branches inside the unit cells. The RMS height and correlation length of the soil roughness were 1.03 and 18.5 cm, respectively. Based on soybean plant spacing of 7.68 cm in-row and 76.5 cm across-rows, a 2x3 unit cell with corresponding Bragg angles of $\varphi = 47.88°$ and $\theta = 44.08°$ was selected to approximate target incidence angles of $\varphi = 45°$ and $\theta = 40°$.

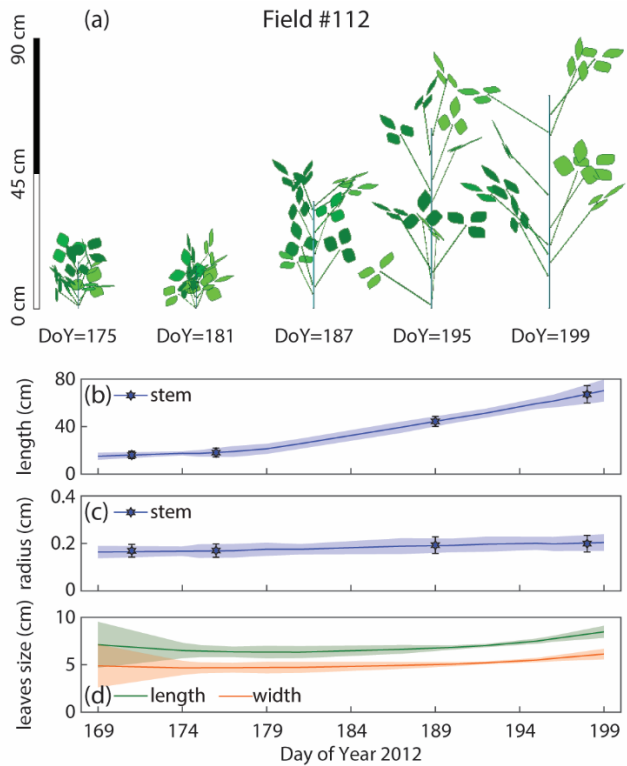

**Fig. 7. The generated 3-D structural models of soybean for field #112.** (a) Sample 3-D structural models generated for 5 days for field #112 are displayed here. (b) The evolution of the stems' length over days is displayed. The shaded area shows the STD. The markers show the measurements, with their STD shown as error bars. (c) The same as panel (b) but for radii of stems. (d) The same as panel (b) but for lengths and widths of leaves.

The average of $\sigma^0$ values from HFSS were compared with the average of $\sigma^0$ values from field observations over a given period using the root mean square difference (RMSD) and STD over 20 realizations. Note that modeled $\sigma^0$ estimates outside $\pm 2$ STD from the observed ones implied that the estimates are significantly (by 95%) different from observations. To demonstrate the confidence interval, the figures show the standard error (SE) of the modeled $\sigma^0$. However, STD for the observed $\sigma^0$ is used in the figures, showing the variability in the observations.

### D. Sensitivity Analysis

The HFSS simulations provide dynamics of $\sigma^0$ and its relationship to changes in crop structure and moisture content throughout the growing season, as well as soil characteristics.

#### D-1) Soil Roughness Analysis

To investigate the impact of soil roughness on $\sigma^0$ [12] [24] [50] [24], we conducted 50 simulations for bare soil, where each simulation employed a unique roughness generated from an RMS height of 1.0 cm, and a correlation length of 12.5 cm. While the generated roughnesses exhibited distinct surface height variations, as shown in Fig 8, they all maintained identical RMS height and correlation length values. The VSM parameter was set to 0.15 $m^3/m^3$ across all simulations. The remaining parameters were made consistent with the SMAPVEX16-MicroWEX simulations.

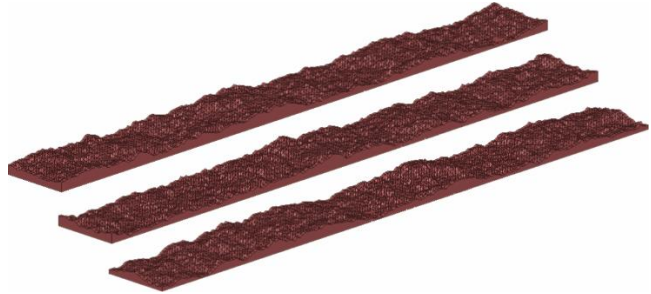

**Fig. 8. Examples of three (out of 50) rough surfaces generated using the same RMS height of 1.0 cm and correlation length of 12.5 cm.** The surfaces exhibit different surface height variations.

*D-2) Sensitivity of $\sigma^0$ to Vegetation Components*

A sensitivity analysis was performed using HFSS to quantify the contributions of individual vegetation components, i.e., stems, branches, leaves, and pods, to the overall $\sigma^0$ signature throughout the growing season. Simulations were conducted with a smooth soil surface and soil moisture of 0.15 $m^3/m^3$, reflecting the mean observed VSM during SMAPVEX16-MicroWEX. These simulations were performed for 9 days spanning the growing season: DoYs 170, 185, 200, 215, 230, 245, 260, 275, and 290, corresponding to the SMAPVEX16-MicroWEX campaign. Eight scenarios were investigated in sensitivity analyses which were modifications of a *base* scenario mimicking the growth of soybean. In the base scenario, the 3-D structural models of soybeans consisted of one stem, ten branches, and one trifoliate per branch (as shown in Fig. 9a). Based upon the SMAPVEX16-MicroWEX observations, the lengths of stems and branches, and the length and width of leaves increased linearly from DoY 170 to DoY 260 with slopes of 1.44, 0.39, 0.08, and 0.03 $cm.day^{-1}$, respectively, and then remained constant till the end of the growing season. The top/bottom radii of stems decreased/increased linearly from DoY 170 to DoY 260 with slopes of 0.01 $cm.day^{-1}$, and increased/decreased by the same slope. The radii of branches were kept constant across the growing season. Additionally, the $M_g$ values of stems, branches, and leaves were monotonically decreasing with the same slope of 0.002 $day^{-1}$. The dielectric properties of vegetation components were estimated from the corresponding $M_g$ values.

In addition to the base scenario, eight scenarios were included: (1) *Fixed-$M_g$*: This scenario presented the sensitivity of $\sigma^0$ to geometry. The $M_g$ of all stems, branches, and leaves were held constant for all 9 days of the simulations, while the geometry was the same as for the base scenario (as shown in Fig. 9b). The constant $M_g$ value was the midseason value observed on DoY 230. (2) *Fixed-geometry*: This scenario explored the sensitivity of $\sigma^0$ to $M_g$. So, while the sizes of vegetation components across the growing season were kept constant at the values corresponding to DoY 230, the $M_g$ of vegetation components varied as in the base scenario (as shown in Fig. 9c). (3) *Stem-only*: This scenario examined the sensitivity of $\sigma^0$ to stems. As a result, branches and leaves were removed from the structural models of the HFSS simulations corresponding to the fixed-$M_g$ scenario, and they were re-executed (as shown in Fig. 9d). The fixed-$M_g$ scenario was selected instead of the base scenario to completely nullify any potential impact of changes in $M_g$ on the outcomes. (4) *No-stem*: To assess the sensitivity of $\sigma^0$ to the absence of stems, a no-stem scenario was implemented, complementing the stem-only scenario. In this scenario, stems were removed from the structural models of the fixed-$M_g$ simulations, and the simulations were re-run (as shown in Fig. 9e). (5) *Branch-only*: The same as the stem-only scenario, but for branches (as shown in Fig. 9f). (6) *No-branch*: the same as the no-stem scenario, but for the lack of branches (as shown in Fig. 9g). (7) *Leaf-only*: The same as the stem-only scenario, but for leaves (as shown in Fig. 9h). (8) *No-leaf*: the same as the no-stem scenario, but for the lack of leaves (as shown in Fig. 9i). To assess the sensitivity of $\sigma^0$ to the presence of pods, a *pod scenario* was simulated. This involved adding pods to the 3-D structural models of the fixed-$M_g$ simulations at DoY 260, corresponding to the peak of the productive stage. The pod clusters were implemented following the same protocol as in the SMAPVEX16-MicroWEX simulations, with lengths and widths matching those observed at DoY 236 of the aforementioned experiment.

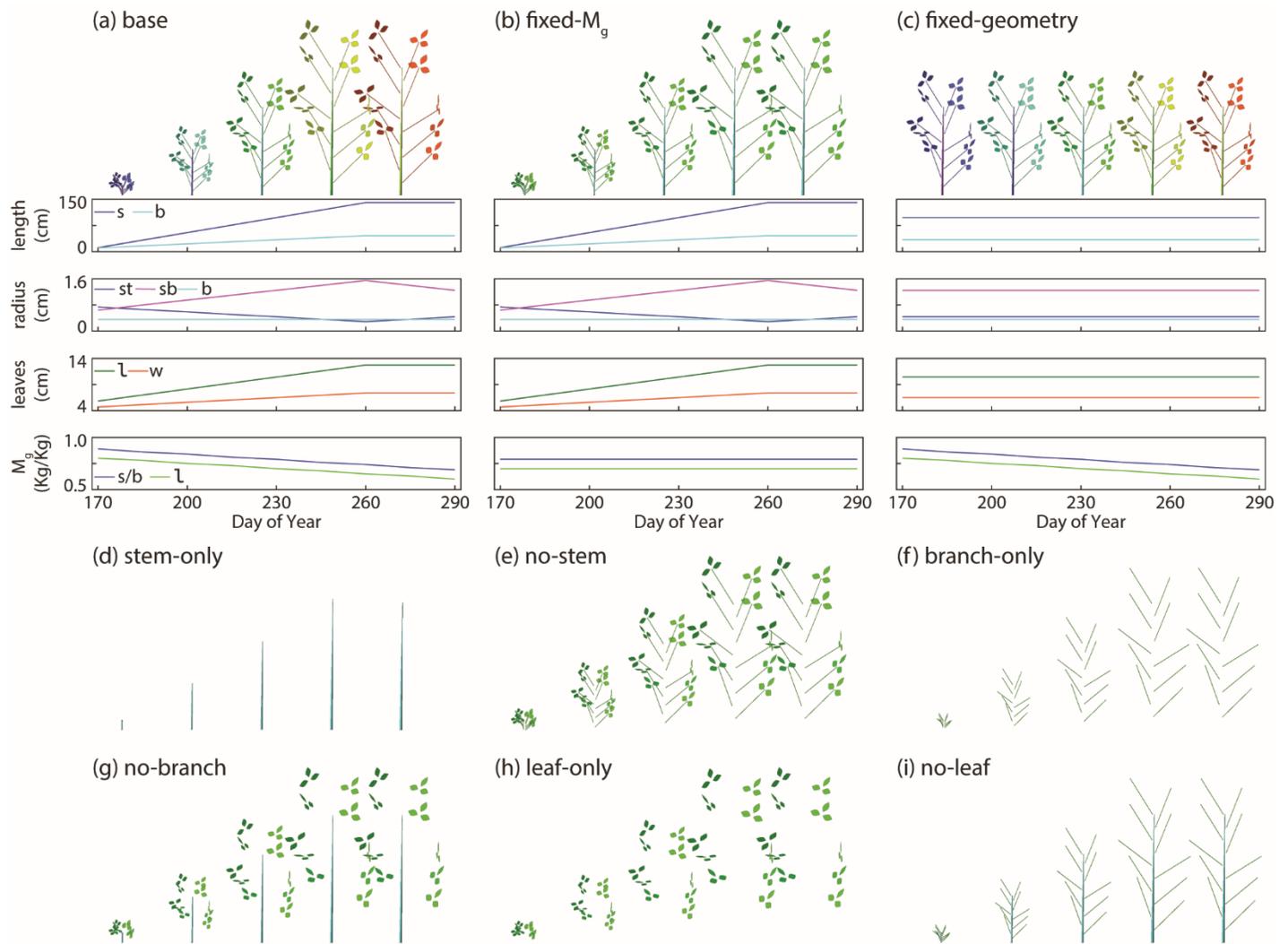

**Fig. 9. The base and eight derived scenarios used for the sensitivity analysis.** (a) This panel, from top to bottom, illustrates the changes in the 3-D structural models of soybean over the growing season, the evolution of the stems' and branches' length, the evolution of the stems' and branches' radius, the evolution of the leaves' length and width, and the evolution of $M_g$ for stems/branches and leaves, under the base scenario. The changes in $M_g$ are visualized by changes in color in the top subpanel. (b) The same as panel (a) but under the fixed-$M_g$ scenario. (c) The same as panel (a) but under the fixed-geometry scenario. (d-i) This panel displays the changes in the 3-D structural models of soybean over the growing season under the scenario mentioned in the panel title.

The distances between synthetic soybean plants along and across rows were set to 8.33 cm and 76 cm, respectively, similar to those for the SMAPVEX16-MicroWEX simulation, and a unit cell size of 2x2 was considered to generate Bragg angles at φ = 0° and θ = 39.13° as the illuminating angles. Simulations were conducted at a frequency of 1.26 GHz. Twenty realizations were generated, each of which was used for the base and eight scenarios. The σ⁰ values were compared with those from the *bare-soil* scenario, which served as the null scenario.

III. Results

*A. HFSS Simulations: SMAPVEX16-MicroWEX*

Fig. 10 shows the HFSS $\sigma^0$ compared with the observations for five days in the early season with DoYs 190-194. During these days, the soybean plants were about 44 cm having about ~11 branches. The average $M_g$ values of stems/branches and leaves were about 0.86 and 0.81 Kg/Kg, respectively (Fig. 10a). During the first three days, the soil moisture decreased by 0.4 $m^3/m^3$, but increased to 0.30 $m^3/m^3$ (Fig. 10b) due to rainfall.

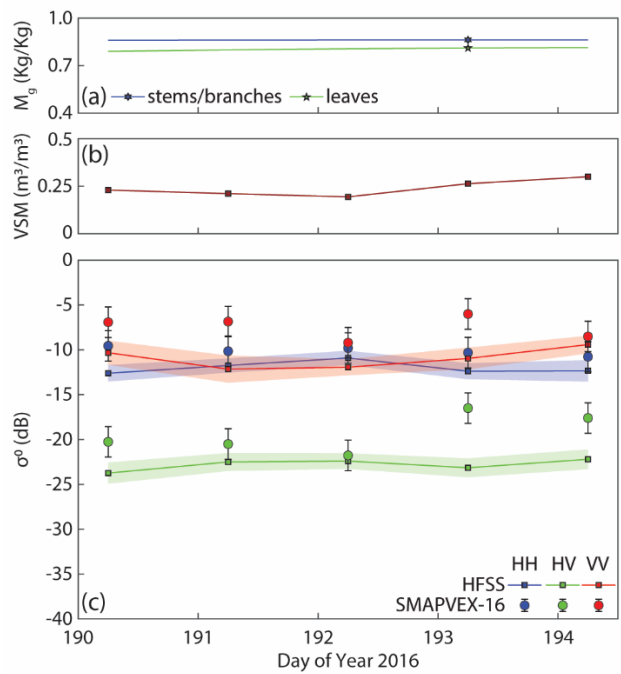

**Fig. 10. The $\sigma^0$ values reported in SMAPVEX16-MicroWEX versus the estimated ones from HFSS over the early days of the growing season.** (a) The evolution of $M_g$ for stems/branches and leaves over the early days is shown here. The markers display the sampling values. (b) The same as panel (a) but for VSM. (c) The measured $\sigma^0$ values (circles: error bars indicate the measurement STD) are compared with the estimated ones from HFSS (squares) at different polarizations over the early days. The shaded area indicates the standard error of the HFSS simulation results.

The estimated $\sigma^0$ values match well with the observations, with the RMSDs of 1.56, 3.82, and 3.04 dB for HH, HV, and VV polarizations, respectively (Fig. 10c). HFSS $\sigma^0_{VV}$ is higher than $\sigma^0_{HH}$, consistent with observations, though the difference between simulation and measurement is lower by 1.44 dB compared to the observations. HFSS $\sigma^0_{HV}$'s are slightly lower than the observed values, except on DOY 193 when the underestimation is 6.83 dB. This deviation could be attributed to unaccounted parameters in the model, such as windy conditions. The averages of modeled estimates were within the ±2 STD of observations.

Fig. 11 depicts the comparison between HFSS estimated and observed $\sigma^0$ values for five days in the late-season with DoYs 246-250. During this period, the soybean plants reached an average height of 109 cm with approximately 12 branches and 43 pods/plant. The $M_g$ of stems/branches and leaves were around 0.77 and 0.70 Kg/Kg, respectively (Fig. 11a). Due to the absence of rainfall, the soil dried steadily, with VSM values decreasing by 0.05 m³/m³ during this period (Fig. 11b).

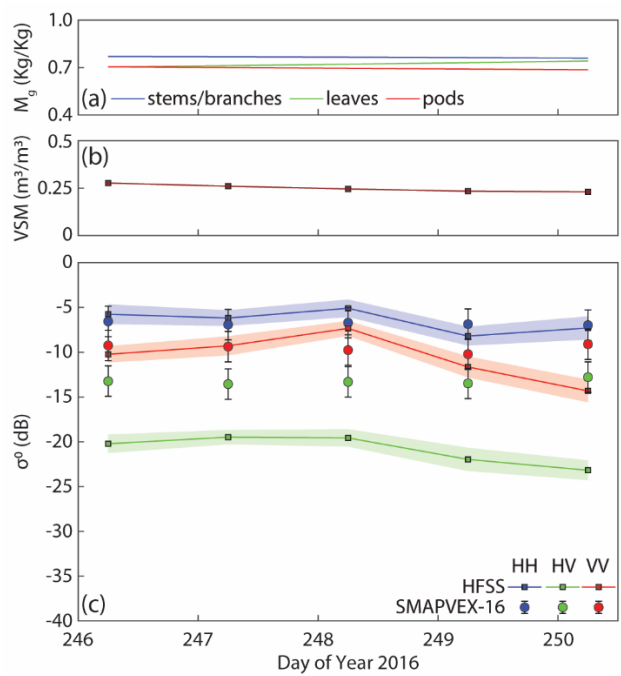

**Fig. 11. The $\sigma^0$ values reported in SMAPVEX16-MicroWEX versus the estimated ones from HFSS over the late days of the growing season.** (a) The evolution of $M_g$ for stems/branches, leaves, and pods over the late days is shown here. (b) The same as panel (a) but for VSM. The markers display the sampling values. (c) The measured $\sigma^0$ values (circles: error bars indicate the measurement STD) are compared with the estimated ones from HFSS (squares) at different polarizations over the late days. The shaded area indicates the standard error of the HFSS simulation results.

The estimated $\sigma^0$ values align well with the measured ones at HH and VV polarizations, with RMSDs of 1.05 and 0.95 dB, respectively (Fig. 11c). However, the estimations at HV polarization are significantly lower than the observations with an average difference of 7 dB (Fig. 11c). Nevertheless, estimations still fall within the ±2 STD band. In both simulations and observations, $\sigma^0_{HH}$ is higher than $\sigma^0_{VV}$, but the simulated difference exceeds the observed difference by 1.23 dB. Notably, the largest divergence between estimations and observations > ±1 STD occurred at HV polarization on DoYs 246-250, reaching a maximum of 9.60 dB on DOY 250. The substantial underestimation in $\sigma^0_{HV}$ is most likely attributed to vegetation structure [10] [51] [14] [52]. It is possible that the branch structure was underrepresented, as information regarding subbranch structure was not collected during the experiment. In addition, observations of cross-pol could also be higher due to some co-pol response leaking into the cross-pol channel.

*B. HFSS Simulations: SMAPVEX12*

Fig. 12 compares the HFSS-simulated $\sigma^0$ values with observations obtained over 12 days for field #63 in the SMAPVEX12. During this period, the soybean plants exhibited substantial growth, transitioning from an initial height of 13 cm to a final height of 60 cm. $M_g$ of stems/branches and leaves exhibited a gradual increase, starting from 0.49 and 0.63 Kg/Kg, respectively, in the early days and reaching approximately 0.84 and 0.82 Kg/Kg, respectively, in the later days (Fig. 12a). Concurrently, the soil dried throughout the simulation period, with VSM values ranging from 0.21 to 0.05 m³/m³ (Fig. 12b).

The HFSS estimated $\sigma^0$ values were consistent with the observed values, with RMSDs of 1.97, 2.54, and 1.76 dB for HH, HV, and VV polarizations, respectively (Fig. 12c). The averages of the modeled estimates consistently fall within the ±2 STD range of the observations. While both observations and simulations reveal a higher average $\sigma^0_{VV}$ than $\sigma^0_{HH}$, the observed difference is smaller than the simulated one by 0.11 dB. The largest discrepancies between estimations and observations > ±1 STD are 3.93 dB at HH polarization on DoY 199, and 2.57 and 3.29 dB at HV polarization on, respectively, DoYs 174 and 175. These discrepancies could be attributed to inaccuracies in the assumed structural descriptions of the soybean models. Due to the limited availability of soybean geometry data during SMAPVEX12, the average geometry assigned to plants across simulation days may deviate from real-world conditions.

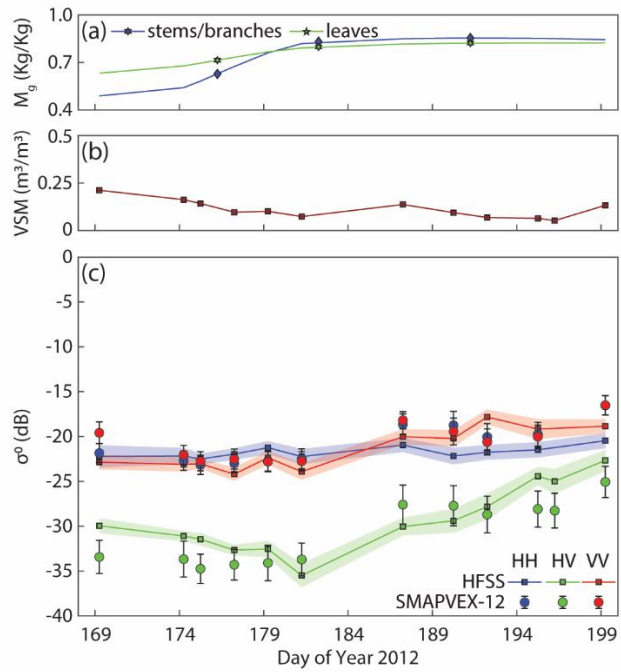

**Fig. 12. The σ⁰ values reported in SMAPVEX12 versus the estimated ones from HFSS for field #63.** (a) The evolution of $M_g$ for stems/branches and leaves over days is presented here. The markers display the sampling values. (b) The same as panel (a) but for VSM. (c) The measured σ⁰ values (circles: error bars indicate the measurement STD) are compared with the estimated ones from HFSS (squares) at different polarizations over days. The shaded area indicates the standard error of the HFSS simulation results.

Fig. 13 compares the HFSS estimated σ⁰ values with observations collected over 12 days for field #82, encompassing a period of significant soybean growth. During this growth phase, plant height increased from an initial 14 cm to a final 66 cm. Notably, $M_g$ of stems/branches and leaves remained relatively constant, ranging between 0.99 and 0.79 Kg/Kg (Fig. 13a). Similarly, VSM was given stability throughout the simulation period, spanning a range from 0.31 to 0.18 m³/m³ (Fig. 13b). The HFSS estimated σ⁰ values exhibit remarkable alignment with the observed values, with RMSDs of 1.60, 3.25, and 1.79 dB for HH, HV, and VV polarizations, respectively (Fig. 13c). Furthermore, the averages of the modeled estimates consistently fall within the ±2 STD range of the observations. While both simulations and observations demonstrate a higher average σ⁰$_{VV}$ than σ⁰$_{HH}$, the observed difference is smaller than the simulated one by 0.46 dB. Discrepancies between observations and estimations, exceeding the 1 STD band, are 5.57 and 5.57 dB at HV polarization on, respectively, DoYs 192 and 196, and 2.44 and 3.86 dB at VV polarization on, respectively, DoYs 169 and 199. These discrepancies could stem from inaccuracies in the assumed 3-D structural models. As for field #63, the simplified representation of crops may overlook crucial geometric features that influence σ⁰, leading to inaccuracies observed in simulation outcomes.

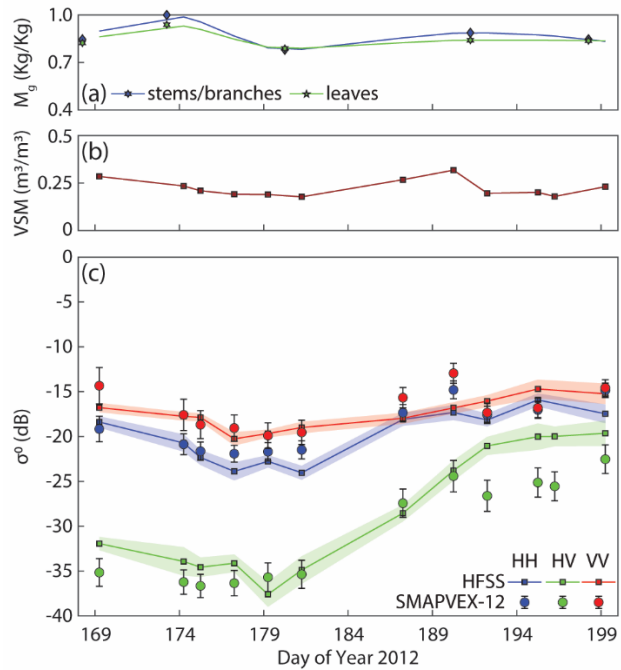

**Fig. 13. The σ⁰ values reported in SMAPVEX12 versus the estimated ones from HFSS for field #82.** (a) The evolution of $M_g$ for stems/branches and leaves over days is presented here. The markers display the sampling values. (b) The same as panel (a) but for VSM. (c) The measured σ⁰ values (circles: error bars indicate the measurement STD) are compared with the estimated ones from HFSS (squares) at different polarizations over days. The shaded area indicates the standard error of the HFSS simulation results.

Fig. 14 presents a comparison between the HFSS estimated σ⁰ values and observations collected over 12 days for field #112, spanning a period of significant soybean growth. During this growth phase, plant height increased from an initial 15 cm to a final 70 cm. $M_g$ of stems/branches and leaves decreased respectively from 0.99 and 0.89 Kg/Kg in the early days to 0.80 and 0.79 Kg/Kg in the later days (Fig. 14a). Similarly, VSM exhibited a steady decline throughout the simulation period, ranging from 0.46 to 0.28 m³/m³ (Fig. 14b). The HFSS estimated σ⁰ values demonstrate notable agreement with the observed values, with RMSDs of 0.93, 2.09, and 1.95 dB for HH, HV, and VV polarizations, respectively (Fig. 14c). Furthermore, the averages of the modeled estimates consistently fall within the ±2 STD range of the observations. While both simulations and observations indicate a higher average σ⁰$_{VV}$ than σ⁰$_{HH}$, the observed difference is smaller than the simulated one by 1.47 dB. Discrepancies between observations and estimations, exceeding the 1 STD band, occurred at VV polarization by 4.13 dB on DoY 192. The deviation could arise from imperfections in the assumed 3-D structural models, as insufficient data on soybean geometry hinders their verification.

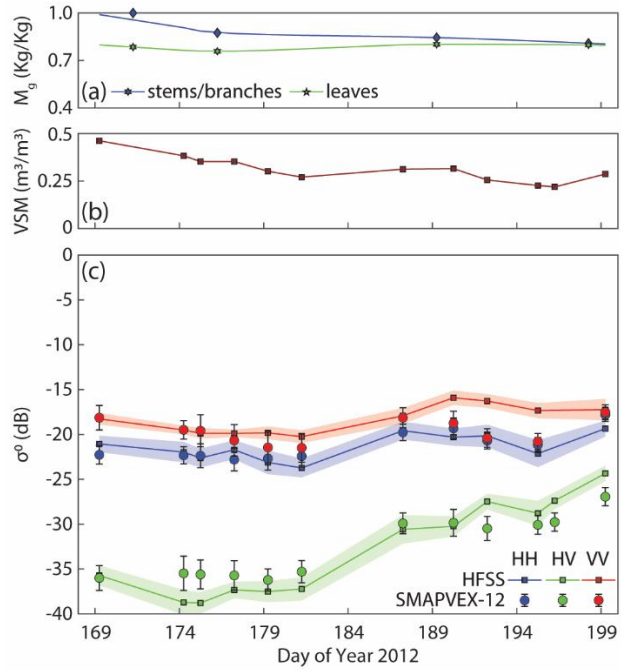

**Fig. 14. The σ⁰ values reported in SMAPVEX12 versus the estimated ones from HFSS for field #112.** (a) The evolution of $M_g$ for stems/branches and leaves over days is presented here. The markers display the sampling values. (b) The same as panel (a) but for VSM. (c) The measured σ⁰ values (circles: error bars indicate the measurement STD) are compared with the estimated ones from HFSS (squares) at different polarizations over days. The shaded area indicates the standard error of the HFSS simulation results.

The three fields experienced almost similar weather conditions, with all undergoing significant soybean growth within the simulation period from DoY 169-199. However, subtle differences existed. For example, field #63 displayed an initial increase in $M_g$ of vegetation components, whereas fields #82 and #112 exhibited an initial decrease. All three fields converged to similar levels of $M_g$ of vegetation components by the end of the simulation period. Soil moisture consistently decreased across all fields, with field #63 being the driest followed by field #82 and then field #112.

The RMSDs of the HFSS estimated σ⁰ values in all fields were consistently below 4 dB across all polarizations. This demonstrates the effectiveness of the HFSS model in capturing the overall backscattering behavior. Field #112 exhibited the lowest RMSDs, suggesting the model performs best under the vegetation and soil parameters reported for this field. However, fields #63 and #82 displayed higher discrepancies between simulations and observations, due to the less accurate representation of the actual soybean structure in the simulations. Despite these inter-field variations, several key similarities emerged. All fields demonstrated a higher average $\sigma^0_{VV}$ compared to $\sigma^0_{HH}$ in both estimations and observations, indicating that the model effectively captures the polarization dependence of $\sigma^0$. Additionally, the averages of the modeled estimates consistently fell within the ±2 STD range of the observations for all three fields, demonstrating the ability of the HFSS model to accurately simulate $\sigma^0$ values under diverse field conditions.

## IV. DISCUSSION

The HFSS model enables the simulation of the evolution of $\sigma^0$ throughout the growing season. Furthermore, it has the capability to demonstrate the variability in $\sigma^0$ due to soil roughnesses, and can illustrate how different vegetation components, such as stems, branches, leaves, and pods, contribute to the overall signature.

### A. The Evolution of $\sigma^0$ Over the Growing Season

The temporal variations of $\sigma^0$ have been extensively studied in various crops, including soybeans [22] [53] [3] [44]. Three distinct phases can be generally identified in the evolution of $\sigma^0$ throughout the growing season: (1) *Ascending phase I*, characterized by vegetative growth leading to canopy closure, during which both $\sigma^0_{VV}$ and $\sigma^0_{HH}$ increase with $\sigma^0_{VV}$ exceeding $\sigma^0_{HH}$. This signature is attributed to the backscattering contribution primarily from soil relative to the crops. (2) *Ascending phase II*, encompassing the period succeeding canopy closure, during which both $\sigma^0_{HH}$ and $\sigma^0_{VV}$ continue to increase; however, due to the higher rate of increase in $\sigma^0_{HH}$ compared to $\sigma^0_{VV}$, $\sigma^0_{HH}$ values exceed those of $\sigma^0_{VV}$ throughout ascending phase II. This shift can be attributed to the increasing contribution of the canopy to backscattering as the season progresses. (3) *The descending phase*, corresponding to the period of water loss and/or shrinkage in the vegetation toward the end of the season, results in decreasing $\sigma^0_{HH}$ and $\sigma^0_{VV}$.

The phased evolution of $\sigma^0$ emphasizes the sensitivity of backscatter characteristics to vegetation growth dynamics and canopy structure, providing valuable insights into the temporal variations of vegetation properties and their impact on $\sigma^0$ behavior. Fig. 15 illustrates the evolution of co- and cross-polarized $\sigma^0$ simulated by HFSS throughout the growing season of SMAPVEX16-MicroWEX. As observed, the HFSS model captures the season's dynamics, with estimated $\sigma^0_{VV}$ and $\sigma^0_{HH}$ exhibiting an increasing trend from DoY 180 to 245, while $\sigma^0_{HH}$ demonstrates a steeper increase compared to $\sigma^0_{VV}$. As a result, while $\sigma^0_{VV}$ dominates $\sigma^0_{HH}$ in the first half of this period, i.e. DoY 180-200, $\sigma^0_{HH}$ surpasses $\sigma^0_{VV}$ in the second half (DoY 200-245). Finally, a decreasing phase is observed from DoY 245 to 250. This confirms that the HFSS model effectively captures the underlying dynamics of the growing season. Soybeans in SMAPVEX16-MicroWEX exhibited significantly accelerated growth compared to those reported in other studies. For instance, while the SMAPVEX16-MicroWEX soybeans attained a height of 1 m at mid-season, the soybeans in [53] required the entire growing season to reach this height. Consequently, the three backscatter phases in SMAPVEX16-MicroWEX are not readily discernible. To address this challenge, a bilinear function was fitted to the model estimations (solid traces in Fig. 15c) to distinguish the backscatter phases. Section IV.C delves into the HFSS model's ability to replicate the phases more efficiently.

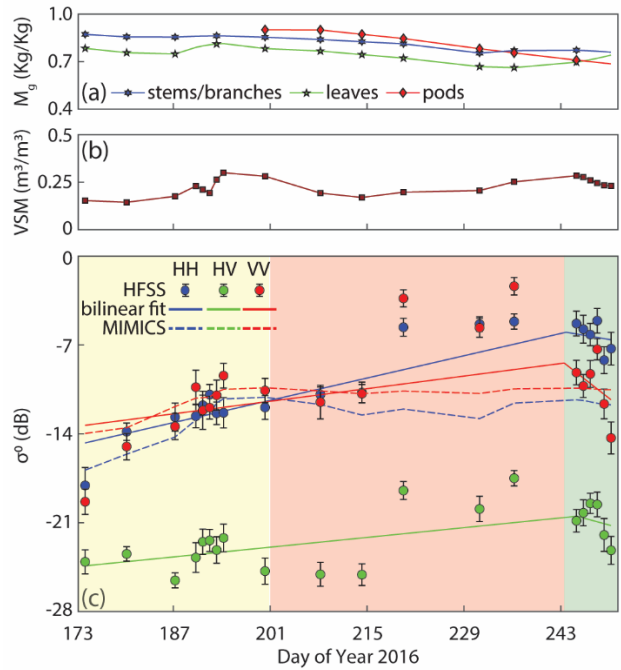

**Fig. 15. The evolution of estimated σ⁰ over the entire growing season.** (a) The temporal evolution of $M_g$ for stems/branches, leaves, and pods throughout the growing season is depicted here. The markers display the sampling values. (b) The same as panel (a) but for VSM. (c) The HFSS estimated σ⁰ values (circles: error bars indicate SE) are compared with the MIMICS estimated ones (dashed) at different polarizations (blue: HH, red: VV, and green: HV). The solid traces display the linear envelopes for the HFSS estimated ones. The yellow, red, and green areas correspond to ascending phase I, ascending phase II, and descending phase, respectively.

Fig. 15 also compares the σ⁰ values from HFSS with the corresponding ones from a widely used Michigan Microwave Canopy Scattering (MIMICS) model [54]. The MIMICS model is based on a first-order solution of the radiative transfer equation for a tree canopy, but it can be applied to estimate the backscatter for various canopy types, including soybean [55] [56]. The MIMICS model's neglect of multiple scattering leads to σ⁰ estimates that are different from HFSS. For instance, while the reported RMSDs for HFSS simulations across diverse regions and growth stages are less than 4 dB (except for one case), MIMICS exhibits RMSDs exceeding 4 dB for co-pol results. Notably, MIMICS lacks the capability to estimate cross-pol σ⁰. Furthermore, representing crop fields with homogeneous vegetation and neglecting row spacing in MIMICS leads to an unrealistic canopy description, particularly during late growth stages. This limitation restricts the estimated evolution of $\sigma^0$ to an initial ascending phase followed by a saturation phase (dashed lines in Fig. 15c). Additionally, MIMICS estimated $\sigma^0_{VV}$ remains always higher than $\sigma^0_{HH}$, indicating an inability to accurately capture the dynamic nature of the problem.

### B. Impact of Soil Roughness on Backscatter

Because HFSS uses 3-D soil structures, it captures the details of soil roughness beyond a two-parameter statistical description used in conventional models such as the Integral Equation Method (IEM) [57]. Thus, HFSS can help provide additional insights into the impact of surface roughness on $\sigma^0$. Fig. 16 compares $\sigma^0$ values obtained from 50 simulations of rough surfaces, each characterized by an RMS height of 1 cm and a correlation length of 12.5 cm. The figure illustrates significant variability in $\sigma^0$ values across these simulations. The interquartile range, representing the middle 50% of the data, spans approximately 8 dB for all polarizations. The medians of the $\sigma^0$ values are -23 dB, -42 dB, and -17 dB for HH, HV, and VV polarizations, respectively. Comparison with Advance Integral Equation Model (AIEM) [58] in Fig. 16 shows that both HH and VV-pol estimates from the AIEM are outside the interquartile range, shown by the colored boxes. AIEM does not provide cross-pol backscatter. This finding necessitates more refined approaches to representing soil roughness in simulations, potentially as distinct scattering centers instead of relying solely on a few statistical parameters. Moreover, it confirms that selecting a realistic roughness is crucial for any full-wave simulation of crop backscatter, since an inappropriate choice of soil roughness can lead to significant deviations in the obtained $\sigma^0$ values.

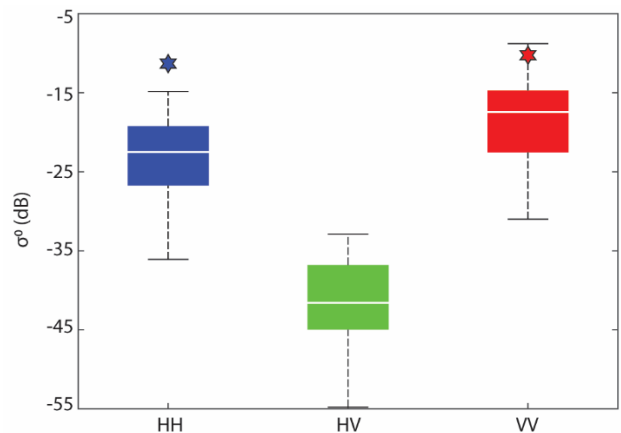

**Fig. 16. The impact of soil roughness on $\sigma^0$ variability.** The variability in HFSS estimated $\sigma^0_{HH}$, $\sigma^0_{HV}$, and $\sigma^0_{VV}$ from 50 soil roughness representations with identical RMS height and correlation length is depicted in blue, green, and red, respectively. The boxes represent the interquartile range and the white line within the boxes represents the median of the data. The hexagram markers represent the Advance Integral Equation Model (AIEM) estimates.

*C. Sensitivity of σ⁰ to Vegetation Components*

The HFSS model was utilized to simulate a growth cycle that mimics the natural development of soybeans in the field (the base scenario). It is important to note that the simulated growth cycle encompassed changes solely in the size and $M_g$ of vegetation components, as illustrated in Fig. 9a. The evolution of $\sigma^0_{HH}$, $\sigma^0_{HV}$, and $\sigma^0_{VV}$ over the growing season is depicted in Fig. 17a. As evident, the proposed growth cycle effectively replicates the ascending phase I from DoY 170-230, ascending phase II from DoY 230-260, and descending phase from DoY 260-290, similar to the simulations from SMAPVEX16-MicroWEX, in Fig. 15.

Fig. 17b shows that the $\sigma^0$ estimates from the fixed-$M_g$ scenario closely resemble the $\sigma^0$ from the base scenario at all polarizations, with the same backscatter phases. This analysis also suggests a minimal influence of $M_g$ on the temporal evolution of $\sigma^0$, and highlights the dominant role of size, or equivalently VWC, in shaping the overall signature. The effect of $M_g$ was investigated by conducting simulations with varying $M_g$ levels while maintaining a constant crop size (the fixed-geometry scenario). In Fig. 17c, the changes in $M_g$ with a fixed geometry produce nearly the same $\sigma^0$ across the growing season without distinguishable backscatter phases, indicating that the $M_g$ has a minimal impact on the overall $\sigma^0$ signature. This analysis implies that crop geometry can be readily estimated from observed $\sigma^0$.

The dominant influence of crop geometry on $\sigma^0$ necessitates further insights into the contributions of individual crop components, namely stems, branches, leaves, and pods to $\sigma^0$ at different polarizations. The simulations corresponding to the fixed-$M_g$ scenario were re-performed with one or more components removed, and the resulting $\sigma^0$ values were compared to those obtained from the fixed-$M_g$ scenario. The fixed-$M_g$ scenario was selected here as the reference instead of the base scenario, given the salient similarity between the base and fixed-$M_g$ $\sigma^0$ (Fig. 17c).

Fig. 17d reveals that the absence of branches and leaves reduces $\sigma^0_{HV}$ to the level of $\sigma^0_{HV}$ for bare soil, and increases $\sigma^0_{VV}$, while $\sigma^0_{HH}$ remains relatively unchanged compared to the fixed-$M_g$ scenario with all the vegetation components. This indicates that the stem is the primary contributor to col-pol $\sigma^0$, but not to $\sigma^0_{HV}$. Additionally, the stem-only scenario confirms the role of branches and/or leaves for the ascending phase II, as the stem-only signature lacks this trend. To corroborate the conclusion drawn from the stem-only scenario, fixed-$M_g$ simulations were conducted without stems, creating a no-stem scenario. Fig. 17e shows that the no-stem $\sigma^0_{HH}$ is close to the $\sigma^0_{HH}$ for bare soil, and no-stem $\sigma^0_{VV}$ is lower than the $\sigma^0_{VV}$ for fixed-$M_g$ with all components, confirming the contribution of stems to co-pol $\sigma^0$. The no-stem $\sigma^0_{HV}$ remains at the level of the fixed-$M_g$ scenario, suggesting that stems do not influence $\sigma^0_{HV}$.

Notably, all ascending/descending phases are absent in the no-stem scenario, implying that stems are essential for the observed three backscatter phases in the growing season.

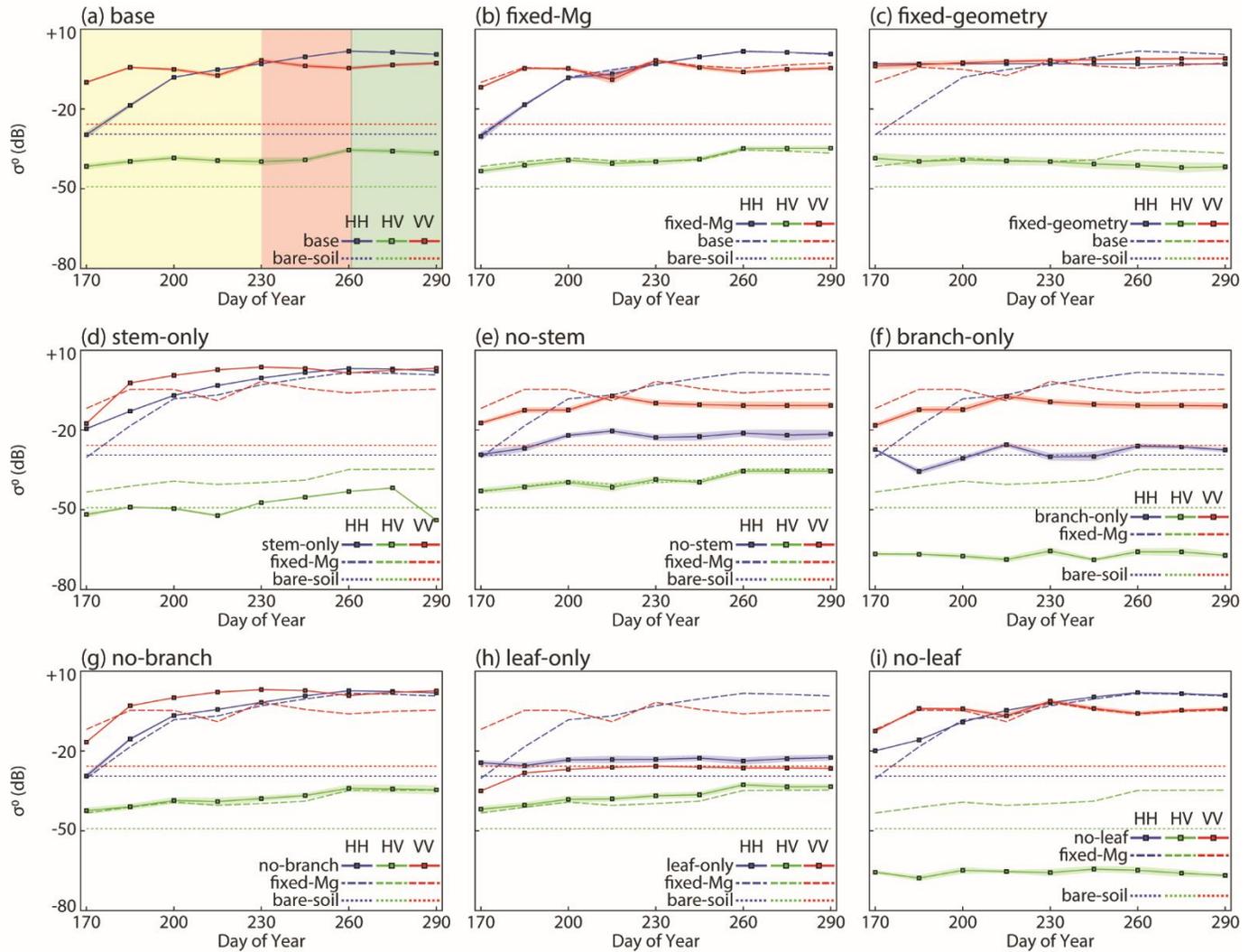

**Fig. 17. The evolution of $\sigma^0$ over the growing season in different scenarios.** (a) The evolution of $\sigma^0$ in the base scenario is shown

here, and compared with the similar one from the bare-soil scenario. The shadings around solid traces represent SE. The yellow, red,

and green areas correspond to ascending phase I, ascending phase II, and descending phase, respectively. (b) The evolution of $\sigma^0$ in the fixed-$M_g$ scenario is shown here, and compared with the similar ones from the base (dashed) and bare-soil (dotted) scenarios. The shadings around solid traces represent SE. (c) the same as panel (b) but for the fixed-geometry scenario instead of the fixed-$M_g$ scenario. (d-i) The evolution of $\sigma^0$ in the scenario mentioned in the title of the panel is compared with the fixed-$M_g$ and bare-soil scenarios.

To isolate the influence of branches on $\sigma^0$, we employed fixed-$M_g$ simulations with stems and leaves removed, creating the branch-only scenario (Fig. 17f). This scenario reduces $\sigma^0_{HH}$ to bare soil levels, demonstrating that branches lack any contribution to $\sigma^0_{HH}$. Interestingly, branch-only $\sigma^0_{VV}$ exhibits a partial reduction compared to fixed-$M_g$, very similar to the no-stem scenario, suggesting a partial contribution of branches for $\sigma^0_{VV}$. Altogether, it appears that the fixed-$M_g$ $\sigma^0_{VV}$ values lay between the $\sigma^0_{VV}$ values observed in the stem-only and branch-only scenarios. Branch-only $\sigma^0_{HV}$ falls below both fixed-$M_g$ and even bare soil, confirming the significant role of branches in reducing cross-pol $\sigma^0$. The absence of backscatter phases reiterates the crucial role of stems in their formation. To validate these findings, we performed the no-branch scenario (Fig. 17g), with only stems and leaves. As expected, $\sigma^0_{HH}$ remains unaffected, confirming the non-contributive role of branches to $\sigma^0_{HH}$. Similarly, $\sigma^0_{HV}$ is unchanged, reiterating the influence of branches in reducing $\sigma^0_{HV}$. However, no-branch $\sigma^0_{VV}$ significantly surpasses fixed-$M_g$ $\sigma^0_{VV}$, closely aligning with the stem-only $\sigma^0_{VV}$. This indicates that the overall $\sigma^0_{VV}$ can be estimated as an average of stem-only and branch-only $\sigma^0_{VV}$ values. Furthermore, the absence of branches eliminates the ascending phase II, as the canopy closure driving this phase no longer exists.

The influence of leaves on $\sigma^0$ was further investigated using leaf-only simulations. As shown in Fig. 17h, leaf-only $\sigma^0$ starkly contrasts with fixed-$M_g$ $\sigma^0$. Both $\sigma^0_{HH}$ and $\sigma^0_{VV}$ are significantly reduced, resembling bare soil. This conclusively demonstrates that leaves do not contribute to co-pol $\sigma^0$. However, $\sigma^0_{HV}$ remains unchanged, indicating that leaves are the primary source of $\sigma^0_{HV}$. Interestingly, the absence of stems and branches leads to the disappearance of all backscatter phases, while $\sigma^0_{HH}$ becomes unexpectedly higher than $\sigma^0_{VV}$. To solidify these findings, the no-leaf scenario was simulated using fixed-$M_g$ with stems and branches only. As depicted in Fig. 17i, the no-leaf $\sigma^0$ closely mirrors fixed-$M_g$ $\sigma^0$ at both HH and VV polarizations, confirming that stems and branches govern co-pol $\sigma^0$. Notably, no-leaf $\sigma^0_{HV}$ is lower than both fixed-$M_g$ and bare-soil $\sigma^0_{HV}$, highlighting the primary role of leaves in $\sigma^0_{HV}$ behavior and suggesting an inhibitory effect of branches on $\sigma^0_{HV}$. These simulations provide compelling evidence that leaves are the primary source of cross-pol $\sigma^0$, while stems and branches primarily contribute to co-pol $\sigma^0$.

This finding contrasts several studies that successfully estimated the Leaf Area Index (LAI) using co-pol $\sigma^0$ [15] [4] [51]. It could be due to a natural correlation existing between leaf size and stem/branch growth, making LAI and stem/branch size dependent variables. While this relationship cannot be investigated experimentally, the HFSS model is able to provide these unique insights. Interestingly, in Section III.A, HFSS simulations for SMAPVEX16-MicroWEX (Fig 12) had underestimated $\sigma^0_{HV}$ compared to observations for late days, DoY 246-250. Through the above analysis, this could be attributed to the underestimation of leaf numbers and/or sizes measured during SMAPVEX16-MicroWEX.

Fig. 18 compares $\sigma^0$ from the vegetative plant in the fixed-$M_g$ scenario for DoY 260 to a reproductive plant on the same day with pods. While the average $\sigma^0_{HV}$ and $\sigma^0_{VV}$ remained relatively unchanged, with p-values of 0.28 and 0.13, respectively, from the Wilcoxon signed-rank test, $\sigma^0_{HH}$ exhibited a significant reduction from 1.78 to -6.95 dB, with a p-value < 0.001. This suggests that pods are indeed contributors to $\sigma^0_{HH}$. Our results partially corroborate [3], where pods exhibit the strongest correlation with the difference between $\sigma^0_{HH}$ and $\sigma^0_{VV}$. However, the analysis in this study reveals that $\sigma^0_{HH}$, not $\sigma^0_{VV}$, is directly correlated with pod presence.

Analyzing the contributions of distinct vegetation components to $\sigma^0$ is crucial in agricultural remote sensing. Elucidating the effects of each component on the overall signature helps improve algorithms for crop type discrimination, as well as crop health and yield estimation. Current models are mainly based on the observed relationship between $\sigma^0$ and biomass/vegetation component [59] [60] [61] [62] [4], and so their applicability is limited to the experiments or regions. In contrast, the HFSS model provides an accurate physically-meaningful estimation of the contribution of each component. Moreover, despite their ability to effectively separate components' effects on $\sigma^0$, current models fall short in accurately assessing the true contribution of each vegetation component due to their limiting assumptions regarding EM wave scatterings. For example, MIMICS restricts wave scatterings to direct, double-bounce, and volume scattering, which may result in unrealistic estimates. While hybrid methods solve Maxwell's equation at the level of individual plants, they neglect multiple scatterings. Consequently, the estimated contribution of each vegetation component using current models is essentially the contribution assigned to that component by the model, and may not be realistic (see, as an example, Pauli decomposition [53]), while the HFSS model presented in this study overcomes these limitations by unrestictedly solving Maxwell's equations.

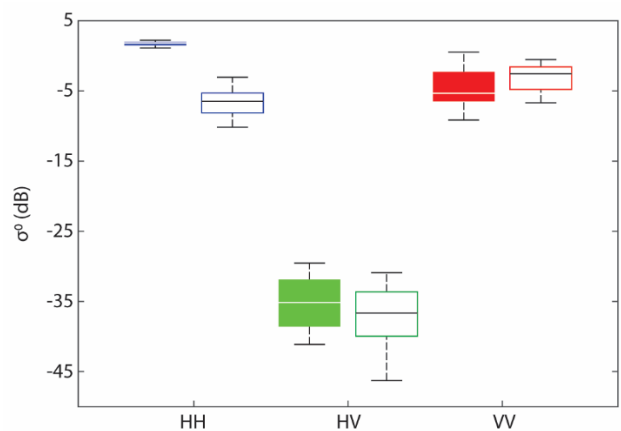

**Fig. 18. The changes in $\sigma^0$ due to pods.** The changes in $\sigma^0$ at different polarizations are depicted by the differences between the colored and empty boxes, which represent $\sigma^0$ values before and after pods, respectively. The boxes represent the interquartile range and the horizontal lines in the boxes show the median of the data.

Overall, the HFSS model offers a substantial leap forward in addressing the challenges of disassociating vegetation component contributions to $\sigma^0$. This ability holds the potential to significantly improve agricultural remote-sensing techniques, where researchers can design retrieval models with improved accuracy. Note that data cubes [63] [64] [65], which are currently the most widely used reverse models and excel at providing a general overview of $\sigma^0$, make it difficult to disentangle the effects of different components and interpret the signature.

## V. CONCLUSION

Full-wave models have emerged as powerful tools for simulating crop backscatter with superior accuracy compared to early and hybrid methods. Unlike traditional models, full-wave models can solve for the scattering of EM waves, particularly backscattering from canopies. This allows them to incorporate factors such as row spacing and detailed 3-D structural models of crops and canopies, which are often neglected in current models.

In this study, we developed a full-wave 3-D model based upon HFSS to demonstrate its potential to accurately estimate $\sigma^0$ across diverse regions and growth stages. During the early season of SMAPVEX16-MicroWEX, the model estimates matched well with field observations with RMSDs less than 4 dB. The same occurred in late-season simulations, with deviations largely confined to cross-pol results, at the level of 7 dB, likely attributable to underestimation of the branch structure or number of leaves in the 3-D structural model, or slightly higher cross-pol measurements due to leakage. Similarly, during SMAPVEX12, the co-pol RMSDs were below 2 dB and cross-pol RMSDs below 4 dB, both during early-season simulations across three fields of observations. Importantly, regardless of the experiment or observation period, modeled estimates consistently fell within the ±2 STD band of observations, highlighting the model's accuracy.

The full-wave 3-D model successfully replicated the evolution of $\sigma^0$ throughout the growing season, including two ascending phases followed by a descending phase. This capability underscores the model's ability to capture the underlying dynamics of the backscattering problem, crucial for identifying key crop development stages and developing more effective inverse models. The bare soil model with a 3-D rough surface demonstrated that roughnesses with the same RMS height and correlation length, the two parameters widely used in scattering models, can induce variability up to 8 dB. Additionally, the HFSS model successfully disassociated the contributions of different vegetation components to the $\sigma^0$ signature, offering valuable insights for developing accurate retrieval algorithms. Our analysis revealed that stems are the primary contributors to $\sigma^0$ at HH polarization, while leaves play a dominant role at HV polarizations. The model also identified branches as co-contributors to $\sigma^0$ at VV polarizations alongside stems, and pods as significant influencers of $\sigma^0$ at HH polarization.

While this study demonstrated some applications of the full-wave 3-D approach, its potential extends beyond the scope of the current research. Future investigations could explore its use with other crops and vegetation types, develop new algorithms for inverse modeling based on this model, and test its performance with various remote sensing data sources. By further refining and applying the model, we can acquire a deeper understanding of the complicated interactions between EM waves and crops, ultimately paving the way for improved crop management practices and enhanced agricultural productivity. The model only serves as a stepping stone towards precise and informative remote sensing models.

### ACKNOWLEDGMENT

The authors would like to thank the computational resources and support provided by the University of Florida High-Performance Computing Center for the simulations conducted in this study.